\documentclass[11pt]{article}

\usepackage[a4paper,lmargin=2.5cm,rmargin=2cm,tmargin=1cm,bmargin=1cm,includehead,includefoot]{geometry}

\usepackage{graphicx,amsmath,url}
\usepackage{microtype,lmodern}
\usepackage{mathtools,amssymb,gensymb}
\usepackage{tgtermes}
\usepackage[T1]{fontenc}
\usepackage[utf8]{inputenc}

\usepackage{acronym}
\usepackage{amsmath}
\usepackage{array}
\usepackage{bbold}
\usepackage{booktabs}
\usepackage{dsfont}
\usepackage{fancyhdr}
\usepackage{relsize}
\usepackage[font={small,up,singlespacing}]{subcaption}
\usepackage[backend=bibtex,style=chem-angew,sorting=none,autocite=superscript]{biblatex}
\usepackage{pdfpages}
\usepackage{bbm}
\usepackage{color}
\usepackage{siunitx}
\usepackage{braket}
\usepackage{mwe}
\usepackage{biblatex}
\usepackage{multirow}
\usepackage{mathtools}
\usepackage{xcolor}
\usepackage{float}
\usepackage{colortbl}
\usepackage{siunitx}\DeclareSIUnit\molar{\textsc{M}}

\usepackage[colorlinks,breaklinks]{hyperref} 
\usepackage{cleveref}

\DeclareMathOperator*{\argsort}{argsort}
\DeclareMathOperator*{\argmin}{argmin}

\title{\textbf{Energy-Based Clustering: Fast and Robust Clustering of Data with Known Likelihood Functions}}

\author{Moritz Th\"urlemann,$^a$ Sereina Riniker$^a$*}
\date{$^a$~Department of Chemistry and Applied Biosciences, ETH Zurich, Vladimir-Prelog-Weg 2, 8093 Zurich, Switzerland. *Email: sriniker@ethz.ch}

\begin{document}
\maketitle

\section*{Abstract}
Clustering has become an indispensable tool in the presence of increasingly large and complex data sets. Most clustering algorithms depend, either explicitly or implicitly, on the sampled density. However, estimated densities are fragile due to the curse of dimensionality and finite sampling effects, for instance in molecular dynamics simulations. To avoid the dependence on estimated densities, an energy-based clustering (EBC) algorithm based on the Metropolis acceptance criterion is developed in this work. In the proposed formulation, EBC can be considered a generalization of spectral clustering in the limit of large temperatures. Taking the potential energy of a sample explicitly into account alleviates requirements regarding the distribution of the data. In addition, it permits the subsampling of densely sampled regions, which can result in significant speed-ups and sublinear scaling.
The algorithm is validated on a range of test systems including molecular dynamics trajectories of alanine dipeptide and the Trp-cage miniprotein. Our results show that including information about the potential-energy surface can largely decouple clustering from the sampling density. 

\section{Introduction}
Uncovering underlying structures in data is essential in the presence of ever-growing data sets.
Clustering is ideally suited to solve this problem \cite{SurveyClustering1, SurveyClustering2}.
The wide range of applications and elusive nature of cluster detection has resulted in the development of a wide range of algorithms. 
Most of these algorithms fall into a small number of families with density-based \cite{DensityBased}, hierarchical \cite{HierarchicalClustering}, partitional \cite{Partitional} or spectral clustering \cite{TutorialSpectralClustering} representing widely used concepts.
However, despite the apparent diversity, most existing clustering algorithms depend either explicitly or implicitly on the sampling density, which can pose issues for cases with insufficiently sampled data. In addition, existing algorithms generally build on the assumption that the similarity between samples can be described by a similarity measure, which is generally chosen to be a metric \cite{SurveyClustering1, InformationClustering, TSNE, DBSCAN, SpectralClusteringNG}. Thus, symmetry is implicitly built into the model. However, for many systems this assumption is clearly violated, for example for transitions between states with different energy. 

Application-specific requirements of data sets in chemistry and physics have resulted in the development of a number of tailored clustering and visualization algorithms \cite{PESTopology, SketchMap, SittelStock, SAPPHIRE1, TMAP}. In particular, the properties of molecular dynamics (MD) simulations data, i.e., large numbers of samples, high dimensionality, and time continuity, underscore the importance of customized methods. Examples for such methods include SAPPHIRE \cite{SAPPHIRE0, SAPPHIRE1}, SketchMap \cite{SketchMap}, CATBOSS \cite{CATBOSS}, or algorithms based on Gaussian mixtures \cite{InfleCS, GMMClustering}.
The introduction of Markov state models (MSM) \cite{MSM1, MSM2} for kinetic modelling based on MD data triggered its own development of clustering algorithms tailored to the specific needs of MSMs, such as free-energy based clustering based on estimated densities \cite{SittelStock} or common nearest neighbours \cite{Gregor}, and algorithms aimed at uncovering core-sets \cite{CoreSet1, CoreSet2, CoreSet3, CoreSet4}. However, although potential energy values are directly available for MD data, no existing clustering algorithm explicitly incorporates this information. Instead, it is generally implicitly introduced based on the sampling density, which can be problematic due to the aforementioned curse of dimensionality and finite sampling effects.

The energy-based clustering (EBC) algorithm proposed in this work was specifically designed to resolve these limitations in the presence of data for which (1) transition probabilities between states are not necessarily symmetric, (2) the sampled density is not fully converged with respect to the underlying distribution, or (3) the sampled density differs from the underlying distribution.
Focusing on the illustrative example of a system conforming to Boltzmann statistics, we demonstrate how inclusion of information about the underlying potential-energy surface (PES), or more generally the likelihood surface, can resolve the issues of finite sampling and curse of dimensionality. Interestingly, the proposed EBC algorithm can be understood as a generalization of spectral clustering, i.e., clustering based on the spectrum of a matrix \cite{Fiedler, SpectralClusteringNG}, in the limit of large temperatures or on undirected graphs.
While the EBC algorithm can be applied to arbitrary data as long as a log-likelihood function or estimate is available, particular emphasis is placed on data sets from MD simulations, where an exact log-likelihood function, i.e., potential energy, is available.
Even though the likelihood may be estimated if the exact function is not known, the main advantage of the proposed procedure over density-based methods, namely independence of the sampled density, is largely lost in such cases. Intuitively, the algorithm observes the behaviour of random walkers on a PES.
Observing the diffusion on a PES permits a concise description of its features through the extraction of the population of each state on one hand and the connectivity between states, on the other hand. 

\section{Theory}
\subsection{Diffusion Matrix}
Let $x_i$ be a state in a state space $X=\{x_i\} \in \mathbb{R}^n$ and $\mathcal{V}(x)$ a function that maps a potential energy $v_i$ to each state. Then, $\mathcal{V}(x)$ describes a PES $V=\mathcal{V}(X)=\{v_i\} \in \mathbb{R}$. Further, given a metric $\mathcal{D}(x_i, x_j)=d_{ij}$ and a cutoff $r$, the neighbourhood of $x_i$ is defined as $N_i=\{x_j \in X \mid \mathcal{D}(x_i, x_j) < r \} \setminus x_i$.

Consider a random walker moving in such a state space. For the following derivation, it is assumed that the transition probability for the transition $i\rightarrow j$ can be described as a product of three independent events: (i) selection of a state within a neighbourhood, (ii) a distance-dependent factor, and (iii) an energy-dependent contribution.
Then, selection of a state within a neighbourhood is inversely proportional to the cardinality of the neighbourhood as only nearby states are accessible, i.e.,
\begin{equation}
    u_{ij} = \frac{1}{|N_i|} .
\end{equation}
A distance-dependent contribution can be included through a distance-based kernel function $K(.)$,
\begin{equation}
    p_{ij} = K(d_{ij}) ,
\end{equation}
with Gaussians being a common choice \cite{TutorialSpectralClustering}.
$p_{ij}$ effectively controls how much a state is allowed to change within one time step of the random walker.

Describing the transition probability purely as the product $u_{ij}\cdot p_{ij}$ presupposes a flat PES.
In situations where one state is more favourable than the other, this assumption is not justified. 
As a solution, inclusion of an energy-dependent contribution based on the acceptance criterion $q_{ij}$ introduced by Metropolis \cite{Metropolis} is proposed 
\begin{equation}
    q_{ij} = \min \left( 1, \exp\left({\frac{v_i - v_j}{T}}\right)\right) 
\end{equation} 
with absolute temperature $T$. 
We note that for large $T$, the family of spectral clustering algorithms \cite{SpectralClusteringNG, TutorialSpectralClustering} can be recovered. 
As a result, the total transition probability $\pi_{ij}$ is described as the product of the three previous terms,
\begin{equation}
    \pi_{ij}=u_{ij} \cdot p_{ij} \cdot q_{ij} .
\end{equation}

Considering all possible transitions $\pi_{ij}$ of a system permits the definition of a transition matrix as 
\begin{equation}
    A = \begin{cases}
                   \pi_{ij} & \text{for } i\neq j\\
                   1 - \sum_{j \neq i} \pi_{ij} & \text{for } i= j\\
        \end{cases}
\end{equation}
which describes the transition probability between each pair of states, including self-loops. The transition matrix defines the diffusive behaviour of random walkers on the PES. It serves as the core of the proposed EBC algorithm and will be used to group states into clusters.

\subsection{Stationary Distribution}
Within a discrete time frame, the population at each state $i$ at time $\tau$ is described as $s^\tau_i$. 
Given an initial population vector $s^0$, the population can be propagated according to
\begin{equation}
    s^{\tau}=s^0A^\tau
\end{equation}
with $A^\tau$ denoting the $\tau$-th matrix power of the transition matrix $A$, and $s^{\tau}$ the population after $\tau$ steps. The flow $A_{ij}^\tau$ describes the fraction of the population at state $i$ transitioning to state $j$ over a period of $\tau$.
If it exists, $s^{\tau}$ will converge towards its stationary distribution $s^{*}$.\cite{Markov} 
In the limit of large $\tau$, the diffusive behaviour can be related to the spectrum of the transition matrix
\begin{equation}
    s^* = \lim_{\tau\to\infty}s^0A^\tau
\end{equation}

We note that finite $\tau$ or eigenvectors with eigenvalues $<1$, respectively, are on one hand necessary for spectral partitioning, but might also be of interest for certain applications, such as the exploration of temporal features, especially since $\tau$ plays a complementary role to the temperature $T$. While $T$ can be used to weight the importance of the depth of the PES, $\tau$ balances short-term and long-term dynamics. 

\subsection{Cluster Extraction}\label{sec:cluster_extraction}
The stationary distribution provides an overview of important states of the system. However, extracting clusters requires further processing. In the following paragraphs, three possible routes to extract clusters are discussed.

\subsubsection{Spectral Cluster Assignment}
As already noted, the proposed algorithm presents a generalisation of spectral clustering, i.e., cluster assignment based on the spectrum of a matrix \cite{Fiedler}.
Here, a recently proposed algorithm based on a QR decomposition by Damle \textit{et al.} \cite{QRRank}, which builds on earlier work by Zha \textit{et al.} \cite{QRRank0}, is used. In short, the proposed method uses a column-pivoted QR decomposition to find a coordinate system that is aligned with a set of eigenvectors. States can then be assigned to a cluster by finding the largest magnitude entry in the new basis for each state.

\subsubsection{Topological Cluster Assignment}\label{sec:topological_clustering}
In addition to the spectral method, we explore a `topological' approach that draws inspiration from topological data analysis \cite{Mapper1, Mapper2, Mapper3}. 
Due to the connection of $\tau$ and $T$ to time and the relative depth of local minima, both parameters can be used to explore how certain states share the same temporal dynamics. 
Accordingly, clusters can be defined based on the notion of `shared dynamics', i.e., by observing where the majority of the population of a state will end up after $n$ time steps. 
From this viewpoint, a simple cluster assignment can be obtained by considering where the majority of the random walker population of a state flows. Specifically, based on the flow over $\tau$ time steps, `attracting sets' $\sigma$ are defined for each state as the $m$ states to which the largest flow is observed,
\begin{equation}
    \sigma_i = \bigcup_{j=1}^{m}\argsort(A_i^\tau)_j .
\end{equation}
A definition of a cluster $C_a$ follows naturally as the set of states for which 
\begin{equation}
    m = |\bigcup_{i\in C_a}\sigma_i| ,
\end{equation}
i.e., all states that share the same attracting set. In the context of graphs, edges can be introduced for attracting sets that overlap to some degree $\frac{|C_a \cap C_b|}{m} > \kappa$.

In other words, this scheme attempts to extract temporal dynamics by considering sets of states, which govern the dynamics of the system over a given time frame $\tau$. Such an approach might provide a complementary view compared to clustering methods which are focused on structural similarity.
One would expect that for small $\tau$ and $T$, the state space will be partitioned into a relatively large number of clusters surrounding neighbouring local minima. For higher temperatures and longer timescales, these clusters will merge and converge at the global minimum given that all states can be reached from all other states. This process offers a natural termination criterion and may also be used to estimate the number of clusters for the QR algorithm.
Combining this concept with ideas put forward in topological data analysis, a scan over $\tau$ and $T$ may be performed to observe the stability and merging of states under changing conditions \cite{Barcodes}.

\subsubsection{Free-Energy Hierarchical Cluster Assignment}\label{sec:hierarchical}
Using the stationary distribution, a free energy may be defined for each state complementary to the equilibrium population as $G_i = -Tln(s_i^*)$
where $s_i^*$ refers to the equilibrium population at state $i$.
Following previous work by Sittel {et al.} \cite{SittelStock} and Weiss \textit{et al.} \cite{Gregor}, the free energy and/or population induces a hierarchy, which can be exploited for cluster formation on one hand and removal of sparsely populated outliers on the other hand.
Specifically, we follow the definition of a cluster used in Ref.~\cite{SittelStock},
\begin{equation}
    S = \{x_i|\mathcal{D}(x_i, x_j) \leq r_{\text{lump}} \land x_j \in S \cap S_0\} .
\end{equation}
A cluster $S$ is formed if there is at least on neighbour $x_j$ within a distance of $r_{\text{lump}}$ in the same set for all states $x_i$ of the considered set, i.e., a connected component. This procedure is repeated for a series of free-energy cutoffs $r_G$ while only states for which $G_i^* < r_G$ are considered at each step.
This approach builds on the intuition that for small free-energy cutoffs only the most highly populated states survive, forming a small number of disconnected clusters. During the successive increase of the free-energy cutoff, transition states between the minima become gradually available, connecting the existing clusters. 

\subsection{Advantages and Practical Considerations}
Including the energy in the transition probability offers two distinct advantages: First, densely sampled states can be replaced by neighbourhoods. As a result, clustering does not have to be performed on all states but on a reduced number of representative low-energy neighbourhoods, which will result in significant speed-ups for densely sampled data.
Second, in the presence of a known likelihood function, explicit inclusion of the likelihood may alleviate finite sampling issues.
As such, insufficient sampling can be compensated for. This means that instead of requiring convergence of the sampled states with respect to the underlying distribution, EBC only requires coverage of states with a high likelihood with respect to the underlying distribution.
Evidently, this is of particular interest for applications where the likelihood function is known, such as the potential energy in MD simulations. 
However, even if the likelihood has to be estimated, the subsequent reduction in states might still improve the computational efficiency compared to standard spectral clustering.

\subsection{Proto-Clusters}
Due to the connection between sampling density and likelihood, performing EBC on densely sampled regions will result in a large degree of redundancy.
As a solution, the notion of a proto-cluster is introduced. Specifically, a proto-cluster $P_i$ is defined as 
\begin{equation}
    P_i=\argmin_V(\{x_j \in X \mid \mathcal{D}(x_i, x_j) < r \}) .
\end{equation}
This means that a proto-cluster is a neighbourhood, which is represented by its lowest-energy state or the average of its members. Membership in a cluster is exclusive. The number of states and distance computations can be greatly reduced in this way. This reduction is possible due to the inclusion of information about the PES and can reduce computational cost considerably. In other words, the local point density is replaced with the potential energy of a neighbourhood and each neighbourhood is represented by its lowest-energy state. 

\subsection{Generalization}
In the current work, the formalism is centered around the Boltzmann distribution and its corresponding negative log-likelihood. Replacing the Metropolis acceptance criterion with its generalization proposed by Hastings \cite{Hastings} permits the application to arbitrary distributions.

\section{Methods}

\subsection{Cantor Potential}
We introduce as a new test system the `Cantor potential'. 
This potential is meant to mimic the fractal nature of PES topologies in a simple test system.
In the present case, this potential was built from a Cantor set generated with five iterations \cite{CantorCantor, CantorSmith}.
The potential was uniformly sampled on the interval $[0, 1]$ with a spacing of $10^{-3}$ and was subsequently normalized. The Cantor set implementation from Ref.~\cite{CantorImplementation} was used. No proto-clusters were used. The temperature was set to $1$ in arbitrary units.

\subsection{10-Well Potential}
The 10-well potential was constructed from $10$ radially distributed Gaussians: A central Gaussian is surrounded by three equally spaced Gaussians placed in a circle around the center. Each Gaussian in this circle is again connected to two Gaussians placed in an outermost circle. 
Gaussians with $\sigma = 2.5$ and an amplitude of $-2.5\,k_B$ were used. The amplitude in the first shell and second shell were scaled by $0.9$ and $0.8$, respectively. For the first leg, the first shell was placed at ($10$, $0$) and the second shell at ($\pm7.66$, $16.43$). The other two legs were obtained through a three-fold rotation.
A trajectory of $1'000'000$ steps was generated using Metropolis Monte Carlo sampling at $300$~K and a step size of $0.5$ in arbitrary units. The potential was constructed and sampled with Ensembler \cite{Ensembler}.
Four exemplary cases were considered based on this trajectory: 
\begin{description}
   \item[Case $0$] The complete trajectory consisting of $1'000'000$ frames.
   \item[Case $1$] The sub-sampled trajectory consisting of $1'000$ frames (every $1'000$th frame).
   \item[Case $2$] $10'000$ frames sub-sampled from the first $100'000$ frames.
   \item[Case $3$] The potential sampled uniformly on a grid.
\end{description}
These cases were constructed to explore dense sampling versus sparse sampling (Case $0$ and Case $1$), finite sampling effects (Case $2$), and distributions that are not sampled from a Boltzmann distribution (Case $3$).

\subsection{Alanine Dipeptide}
Alanine dipeptide was simulated with the AMBER ff99SB-ILDN force field \cite{99SB, 99SBILDN} in combination with the GBn2 implicit solvation model \cite{GBn2}. A $1\,\mu$s trajectory was sampled at $350$~K with OpenMM (version 7.7) \cite{OpenMM7}.
Bonds involving hydrogens were contrained with LINCS \cite{LINCS}. The trajectory was sampled using a Langevin integrator with a $2\,$fs time step and a collision frequency of $1\,$ps. No cutoff was applied to nonbonded interactions. The $\phi$ and $\psi$ dihedral angles of the backbone were used as features for the clustering.

In addition, the ETKDG conformation generator \cite{ETKDG} as implemented in the RDKit \cite{RDKIT} was used to generate $10'000$ conformations. No RMSE pruning was applied. To obtain potential energies for the conformers, the same force field and implicit solvent model as in the MD simulation were used. 

\subsection{Trp-cage Folding}
The Trp-cage protein was simulated using the AMBER ff99SB-ILDN force field \cite{99SB, 99SBILDN} in combination with the GBn2 implicit solvation model \cite{GBn2}. A $5\,\mu$s trajectory was simulated with OpenMM (version 7.7) \cite{OpenMM7}.
The folded protein was first unfolded in a $1\,$ns simulation at $500$~K, followed by the folding simulation over $5\,\mu$s at $300$~K. As the reference, the structure resolved by Neidigh \textit{et al.} was used (PDB code: 1L2Y) \cite{TrpCageExp}.
Bonds involving hydrogens were in all cases fixed. The trajectory was sampled using a Langevin integrator with a $2\,$fs time step and a collision frequency of $1\,$ps. No cutoff was applied to nonbonded interactions.
As features for the clustering, the first three principal components of the Cartesian coordinates of the aligned backbones were used, explaining $54\%$ of the variance.

\subsection{Implementation and Hyperparameters}\label{sec:implementation}
The proposed EBC algorithm was implemented with Python (3.9) \cite{Python}, using NumPy (1.19.5) \cite{Numpy}, SciPy (1.7.3) \cite{scipy}, scikit-learn (1.1.1) \cite{SKLearn}, and NetworkX (2.6.3) \cite{Networkx}. The Kamada-Kawai layout \cite{KamadaKawai} as implemented in NetworkX was used for graph visualisation. Matplotlib (3.3.2) \cite{Matplotlib} and seaborn (0.12.1) \cite{Seaborn} were used for plotting. Protein structures were visualised with PyMOL (2.5.2) \cite{Pymol}.
MDTraj was employed to process the sampled MD trajectories \cite{MDTraj}.
For simplicity, all results shown in the following sections use the same hyperparameters and a k-nearest neighbor (KNN)-based implementation. The KNN-based implementation used $8$ nearest neighbours, i.e., $u_{ij} = \frac{1}{8}$ in all cases and $p_{ij} = 1$. Energies were scaled to a standard deviation of $1$ and the temperature was set to $1$. The proto-cluster radius was automatically estimated from the $0.8$ percentile of samples from the distribution of distances between states.

For the Trp-cage MD trajectory, the unscaled potential energies were used. The temperature was set to $T=100$~K and KNN~=~$16$. A fixed proto-radius of $0.5$ was used resulting in $2'482$ proto-clusters for the complete trajectory consisting of $250'000$ frames.
For the topological cluster assignment (see Section \ref{sec:topological_clustering}), attracting sets with $m=8$ were used. Edges were added for an overlap of $\geq \frac{7}{8}$.

\section{Results and Discussion}
In the following paragraphs, results for several test cases and systems are presented to illustrate features of the proposed EBC algorithm. Results are structured into the application to test systems in Section~\ref{sec:cantor},~\ref{sec:cases},~\ref{sec:cost} and the application to MD trajectories of alanine dipeptide (Section ~\ref{sec:diala}) and a folding trajectory of Trp-cage (Section~\ref{sec:trp_cage}). In Appendix~\ref{sec:sklearn}, application to standard datasets available in the scikit-learn package \cite{SKLearn} are provided.

\subsection{Cantor Potential}\label{sec:cantor}
\begin{figure}[H]
    \centering
    \includegraphics[width=0.6\textwidth]{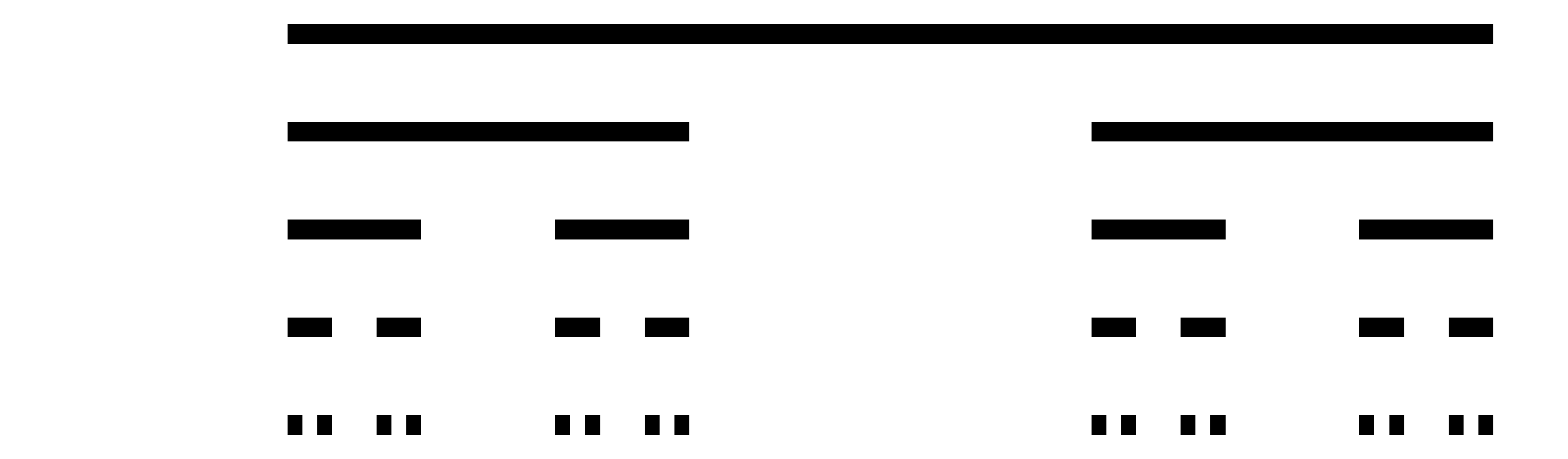}
    \includegraphics[width=0.6\textwidth]{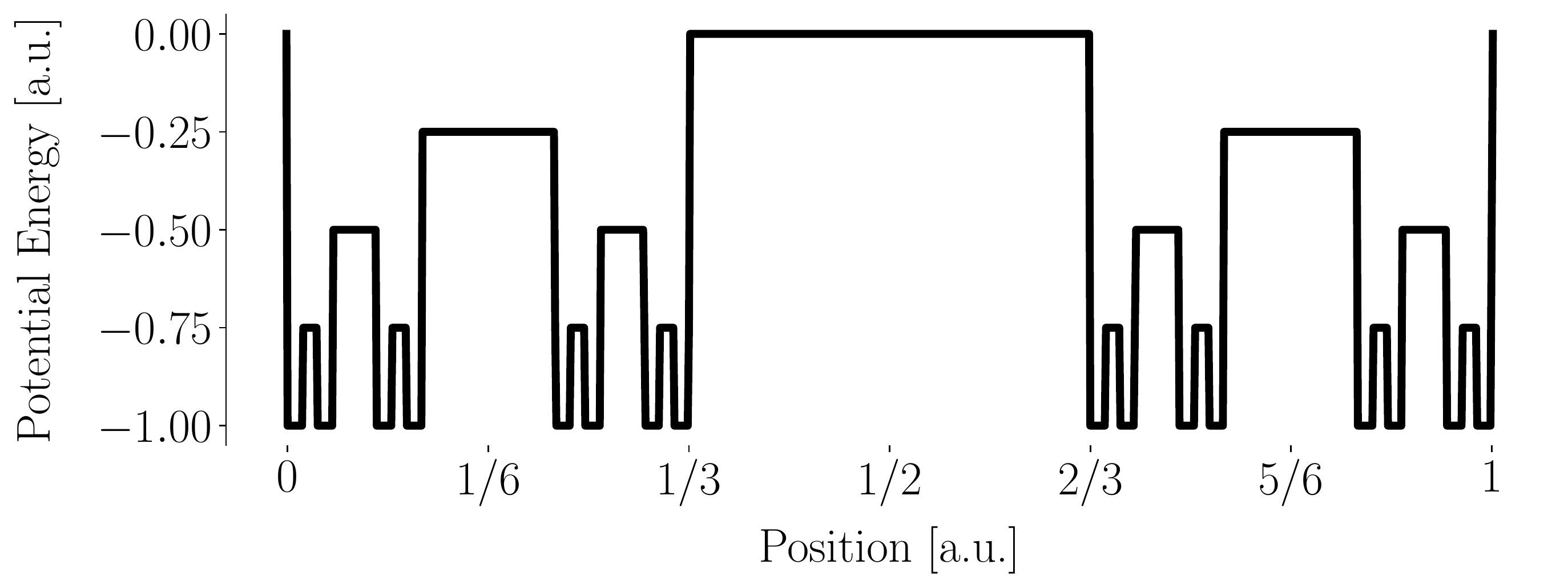}
    \includegraphics[width=0.6\textwidth]{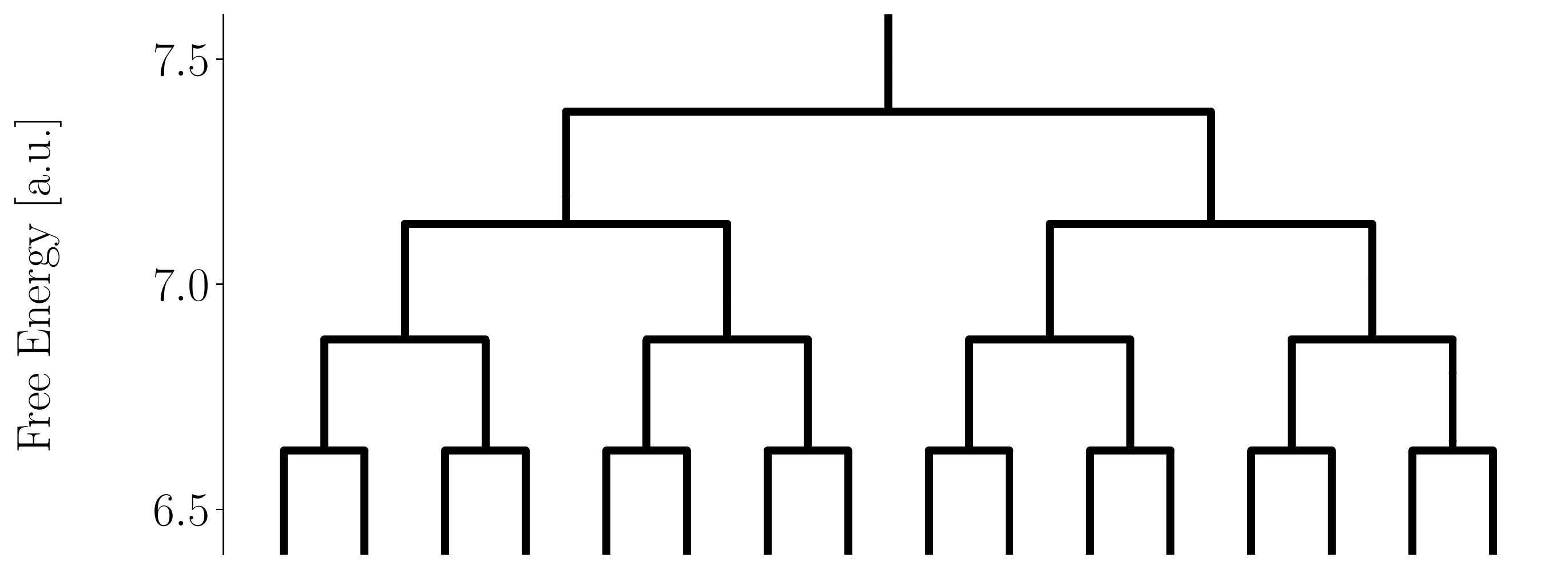}
\caption{(Top): Cantor set based on $5$ iteractions. (Middle): Resulting potential-energy surface (PES). (Bottom): Topology of the PES recovered by the free-energy based hierarchical cluster assignment.}
\label{fig:cantor}
\end{figure}

As a first example, a PES based on the Cantor set is introduced (Figure~\ref{fig:cantor}). 
This case was designed to explore the ability to recover the topology of a PES through free-energy hierarchical cluster assignment discussed in Section~\ref{sec:hierarchical}.
As shown in Figure~\ref{fig:cantor}C, EBC can reliably recover the fractal nature of the topology of this PES. While the reconstruction is only shown for $5$ iterations here, arbitrary numbers of levels are in principle possible if the intervals are sampled sufficiently fine. Besides the detection of transition states, free-energy based hierarchical cluster assignment can be used to inform the number of clusters during a subsequent spectral cluster assignment. 

\subsection{10-Well Potential}\label{sec:cases}
A 10-well potential was designed to mimic finite sampling effects, for example due to kinetic effects, which may result in an unequal sampling density for states with the same likelihood. Four different cases were distinguished.

\begin{figure}[H]
\begin{subfigure}[b]{0.48\textwidth}
    \centering
    \includegraphics[width=0.8\textwidth]{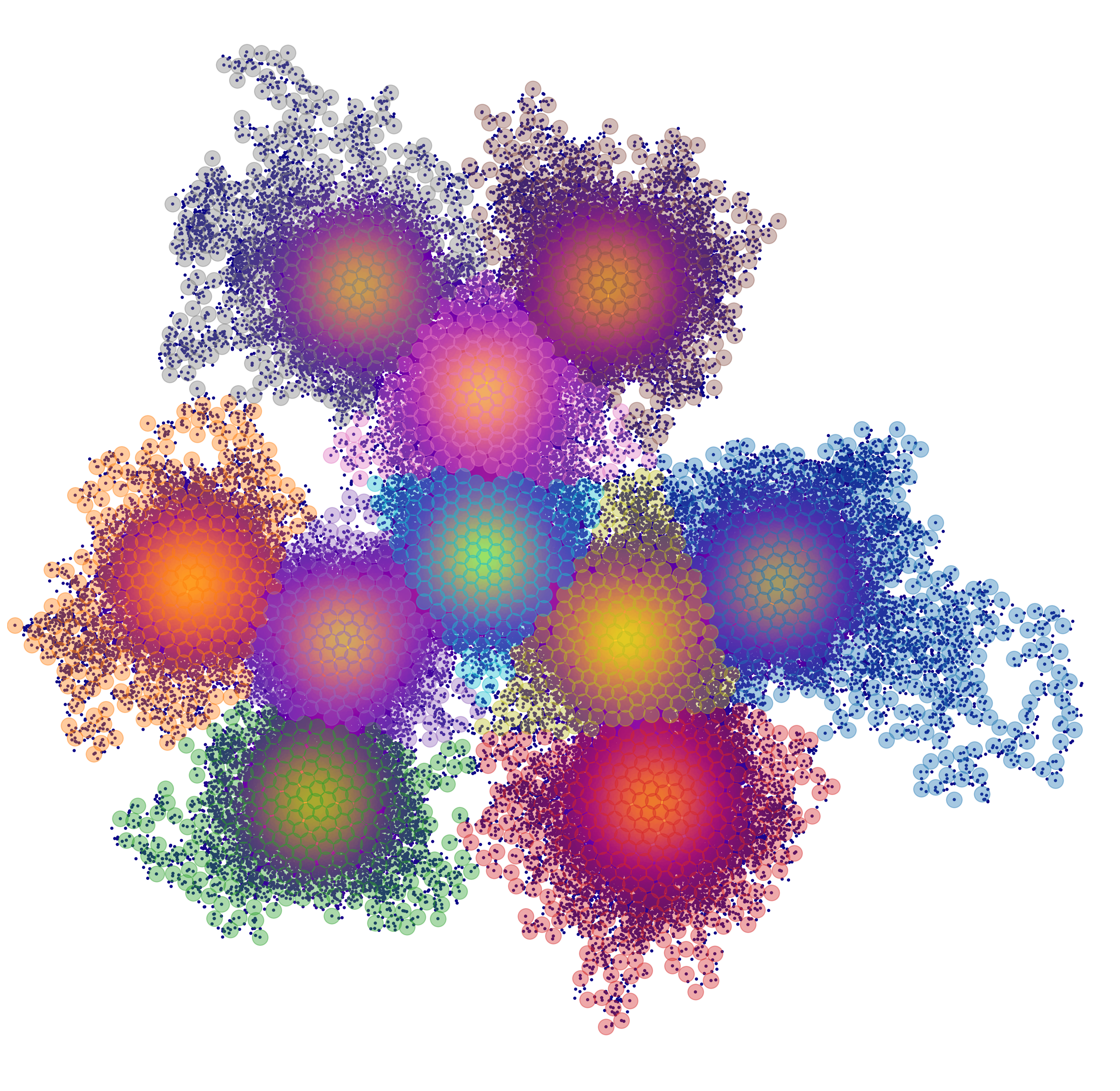}
    \caption{Case $0$}
\end{subfigure}
\begin{subfigure}[b]{0.48\textwidth} 
    \centering
    \includegraphics[width=0.8\textwidth]{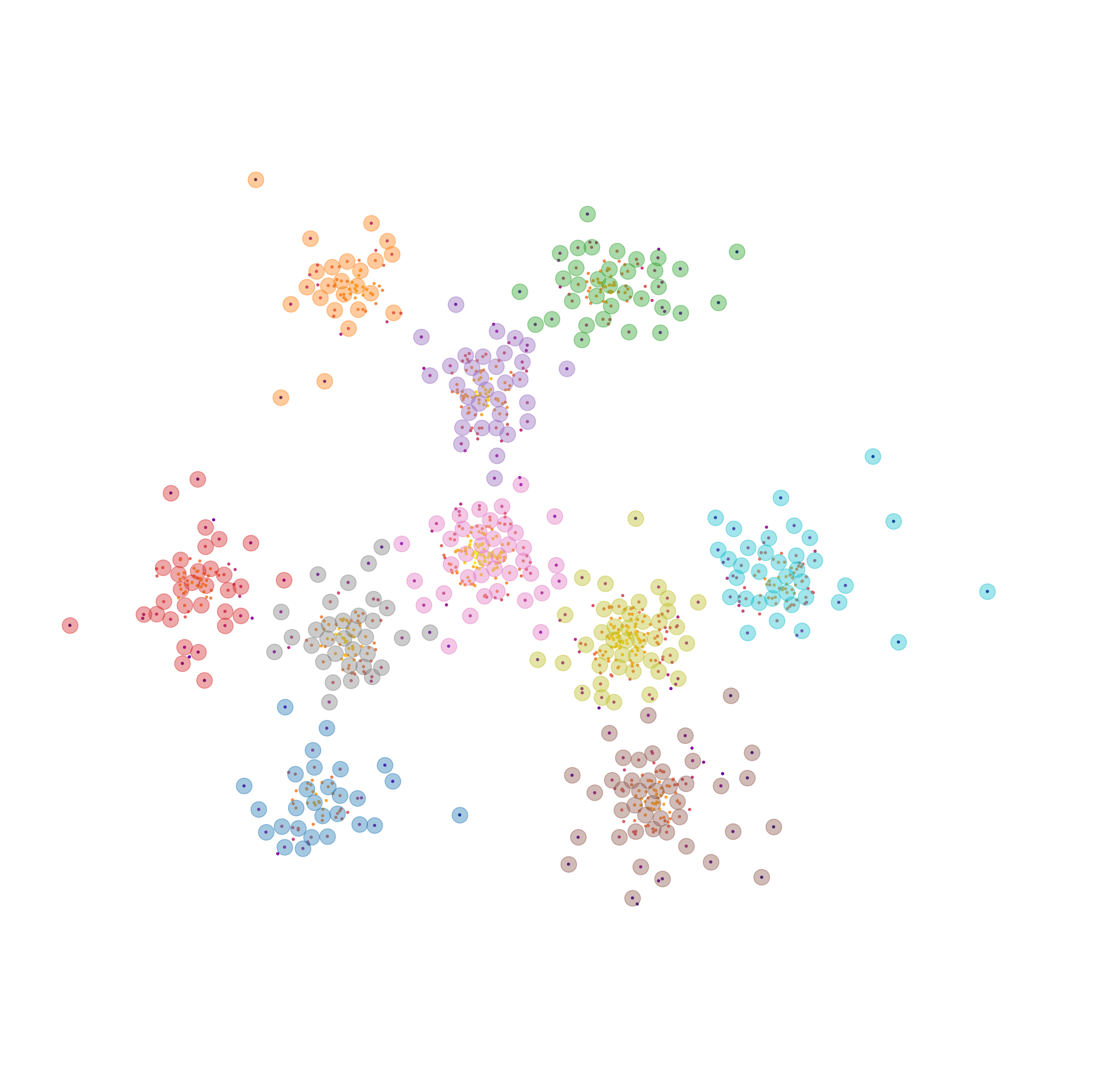}
    \caption{Case $1$}
\end{subfigure}
\begin{subfigure}[b]{0.48\textwidth} 
    \centering
    \includegraphics[width=0.8\textwidth]{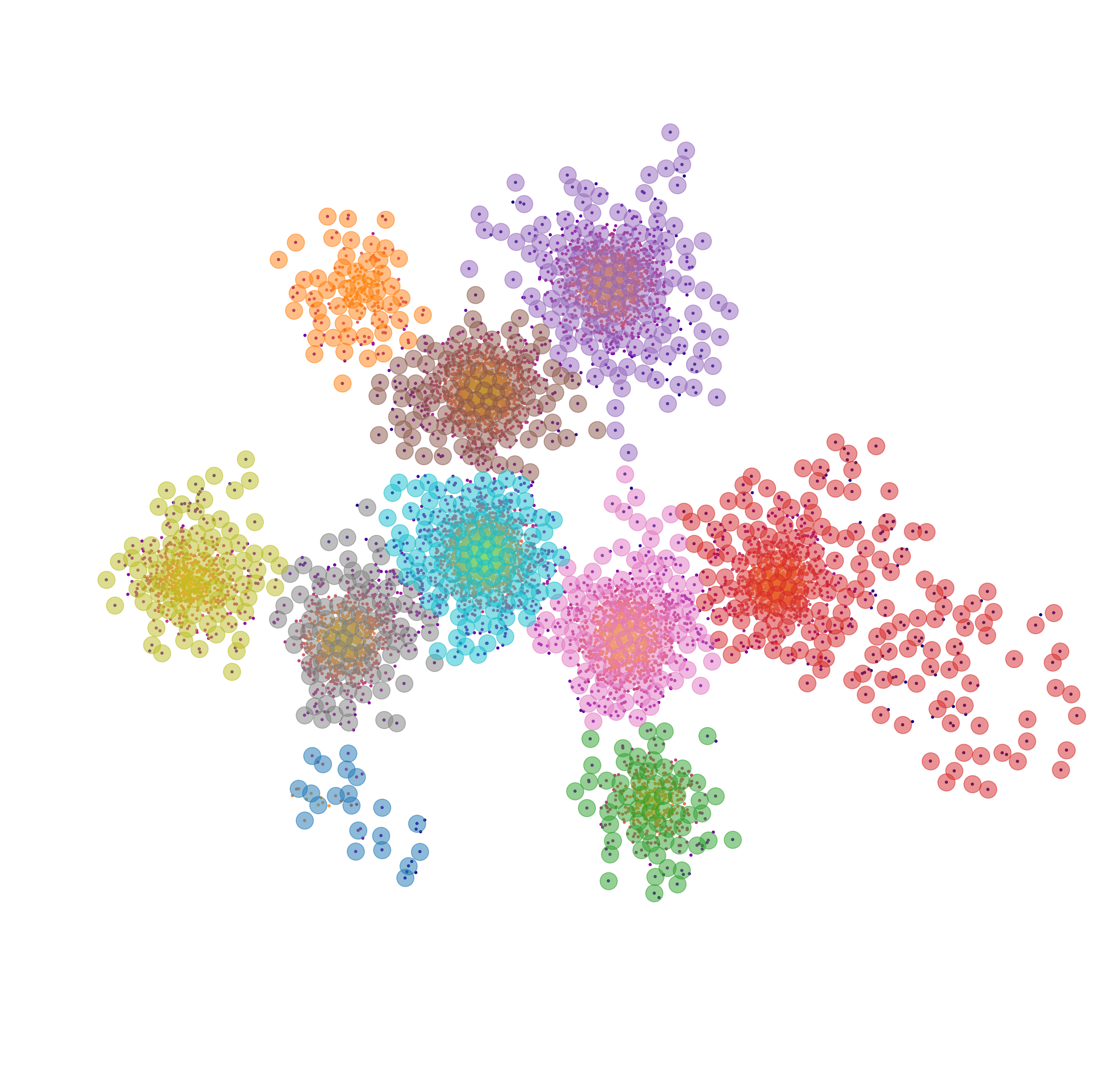}
    \caption{Case $2$}
\end{subfigure}
\begin{subfigure}[b]{0.48\textwidth}
    \centering
    \includegraphics[width=0.8\textwidth]{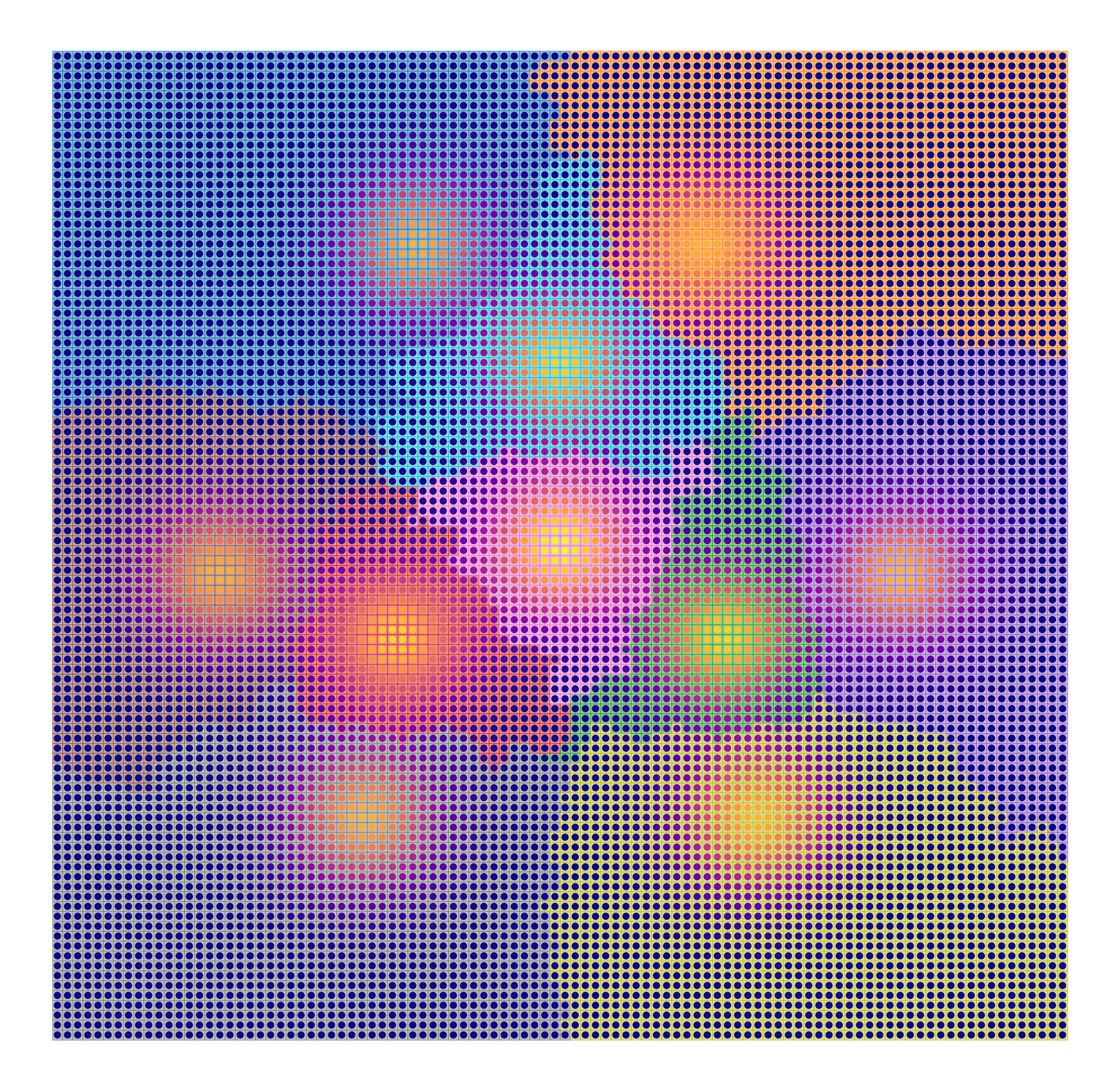}
    \caption{Case $3$}
\end{subfigure}
\caption{EBC results for case $0$ (full trajectory), case $1$ (subsampled trajectory with $1000$ data points), case $2$ (insufficient sampling), and $3$ (different potential uniformly sampled) of the 10-well potential. Small dots in the background show data point, with brighter colors indicating lower potential energies. Large circles represent proto-clusters with the respective cluster membership indicated by the respective color.}
\label{fig:10_well_cases}
\end{figure}

In case $0$, the complete trajectory of $1$ million data points sampled with Metropolis Monte Carlo was used for clustering, whereas in case $1$ only every $1'000$th frame was used, resulting in a total of $1'000$ samples. For the estimated proto-radius of $0.6$ (case 0) and $0.58$ (case 1), a total of $2'753$ and $353$ proto-clusters, respectively, were found.
The same clusters corresponding to the potential minima were identified in both cases (Figure \ref{fig:10_well_cases}).

In addition, two particularly challenging cases were considered: Case $2$ with insufficient sampling and case $3$ consisting of the uniformly sampled potential, which differs from the underlying Boltzmann distribution.
To simulate finite sampling effects (case $2$), only every $10$th frame of the first 10~\% of the trajectory was used. As shown in Figure \ref{fig:10_well_cases}c, not all minima are sampled equally as a result. Particularly the well in the lower left and upper left are sampled much more sparsely than the more central wells. Nevertheless, the algorithm finds the same $10$ wells as with the full trajectory (case $0$ and $1$).

Finally, case $3$ demonstrates the performance of the EBC algorithm for datasets where the sampled density differs from the underlying distribution. For this purpose, the 10-well potential was sampled uniformly on a grid. 
Density-based clustering algorithms fail in such cases.
Figure \ref{fig:10_well_cases}d shows how the clusters at the minima and surrounding regions are still correctly assigned with the EBC algorithm.
No other tested algorithm yielded the same results. When removing high-energy data points (i.e., sampling only around the minima), several algorithms like KMeans++ \cite{KMeans++} and distance-based spectral clustering were able to identify the correct clusters (data not shown).

\subsection{Computational Cost}\label{sec:cost}
As discussed before, inclusion of energies allows for the use of proto-clusters. A proto-cluster subsampling can result in massive speed-ups compared to a clustering of all samples, particularly for densely sampled data. 
Table \ref{tab:timing} lists the execution time required to cluster trajectories of the 10-well potential of increasing size: $1'000$, $10'000$, $100'000$, and $1'000'000$ frames.
The same proto-cluster radius of $0.6$ was used in all cases, resulting in $341$, $959$, $1'944$, and $2'756$ proto-clusters, respectively. Run times were measured on the same standard desktop workstation over $10$ runs and $10$ repetitions each. The values in Table \ref{tab:timing} are given relative to the time obtained for $1'000$ samples, which was measured as $41.4\pm\,1.8$ms. 
Timings for KMeans++ \cite{KMeans++} and DBSCAN \cite{DBSCAN} were measured on the same machine using the implementations available in scikit-learn \cite{SKLearn}.

DBSCAN outperforms the other algorithms for small numbers of data points. However, this advantage is lost for larger numbers. Due to crashes, it was not possible to obtain a measurement for $1$M samples using DBSCAN.
Thanks to the proto-cluster based formulation, the EBC algorithm exhibits sub-linear scaling over three magnitudes. It must be noted that the gain due to the proto-cluster subsampling depends strongly on the density of the data points. In the presence of large numbers of proto-clusters, the eigendecomposition will become the limiting factor. 

\begin{table}[H]
    \centering \small
    \begin{tabular*}{\textwidth}{ccccc}
    \hline
    \multicolumn{5}{c}{Execution Time}\\
    Number of Samples & $1$K & $10$K & $100$K & $1$M\\\hline \hline
    EBC & 1 ($41.4\pm 1.8$ms) & 2.8 ($117\pm 0.6$ms) & 8.3 ($343\pm 4.43$ms) & 57.7 ($2.39\pm 0.05$s) \\\hline
    KMeans & 0.6 ($23.2\pm 4.5$ms) & 22.5 ($932\pm 10.3$ms) & 31.2 ($1.29 \pm 0.01$s) & 111.4 ($4.61\pm 0.03$s)\\\hline
    DBSCAN & 0.1 ($4.49\pm 3.1$ms) & 1.6  ($67.3 \pm 0.4$ms) & 52.4 ($2.17\pm 0.01$s) & - \\\hline
    \end{tabular*}
    \caption{Execution time for clustering of $1'000$, $10'000$, $100'000$ and $1'000'000$ frames of the trajectory of the 10-well potential. KMeans++ and DBSCAN are shown for comparison. Relative execution times are reported. Absolute times $\pm$ one standard deviation are shown in brackets.}
    \label{tab:timing}
\end{table}

\subsection{Alanine Dipeptide}\label{sec:diala}
\begin{figure}[H]
\begin{subfigure}[b]{0.99\textwidth}
    \centering
    \includegraphics[width=0.99\linewidth]{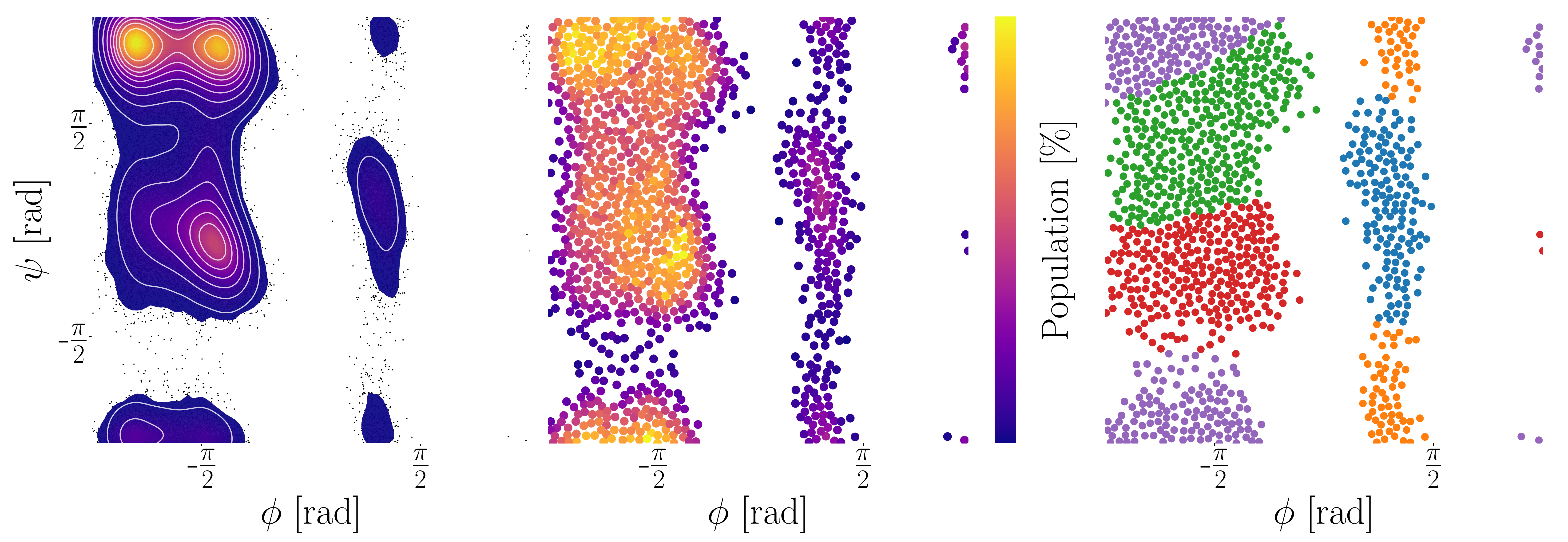}
    \caption{Sampling with MD}
\end{subfigure}
\begin{subfigure}[b]{0.99\textwidth} 
    \centering
    \includegraphics[width=0.99\linewidth]{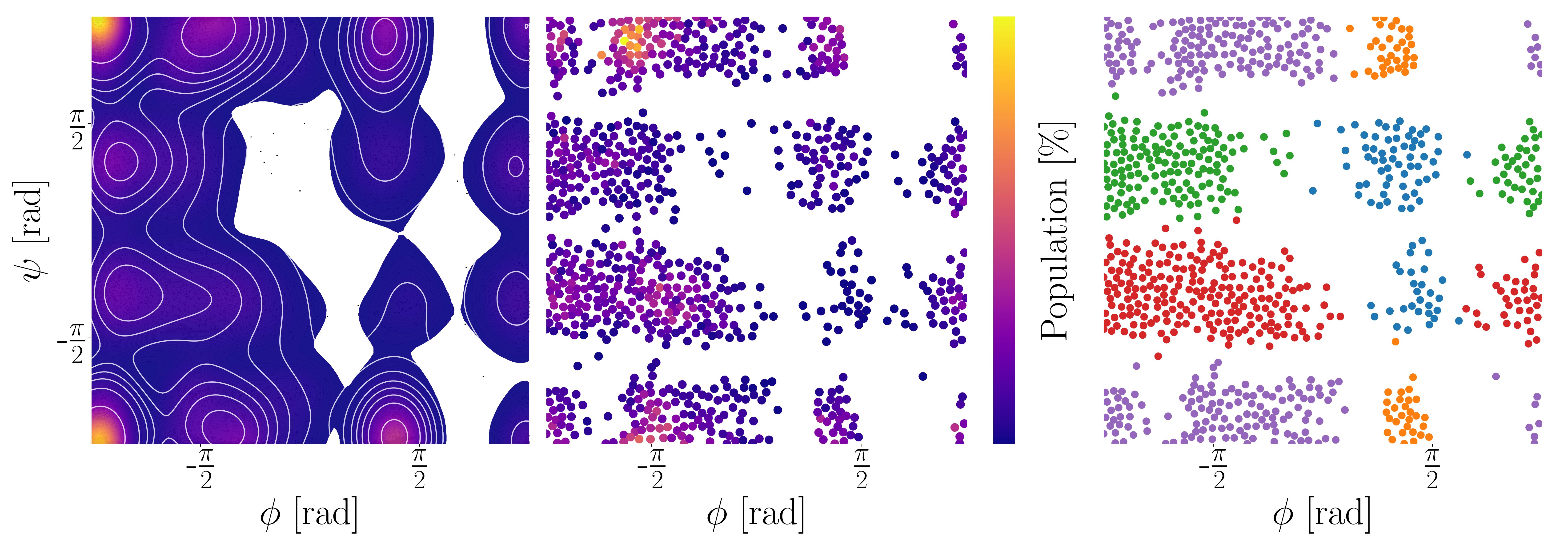}
    \caption{Sampling with ETKDG}
\end{subfigure}
\caption{Conformational ensembles of alanine dipeptides sampled with MD (top) and the ETKDG conformation generator (bottom): sampled point densities (left), population density of the stationary distribution from the EBC clustering (middle), and cluster labels (right).}
\label{fig:dalanine}
\end{figure}

Two sampling approaches were investigated for alanine dipeptide: (i) sampling with MD simulations, and (ii) sampling using an \textit{in silico} conformation generator (ETKDG \cite{ETKDG}). MD offers in principle a simple way to sample extensively from a Boltzmann distribution, however, finite sampling effects and/or high computational costs can limit the effectiveness in practice.
In contrast, \textit{in silico} conformation generators provide an efficient way to generate large sets of conformers, but they rely typically on stochastic or systematic sampling and do not sample from a well-defined distribution.
Conformational ensembles of alanine dipeptide can be visualised and compared using a Ramachandran plot of the backbone dihedral angles (Figure~\ref{fig:dalanine}).

Using MD, the known free-energy landscape for the backbone dihedral angles of alanine dipeptide in implicit solution \cite{DalaFESurface} was recovered (Figure~\ref{fig:dalanine}a).
Even though the distributions from MD and ETKDG differ substantially (compare left panels in Figure~\ref{fig:dalanine}), good qualitative agreement between the resulting clusters is obtained for both cases as shown in the right panels of Figure~\ref{fig:dalanine}.   
Using EBC on conformational ensembles (from MD, Monte Carlo, or conformer generators) could be a valuable approach to obtain weighted ensembles or to diverse starting conformations for subsequent simulations.

\begin{figure}[H]
    \centering
    \includegraphics[width=\textwidth]{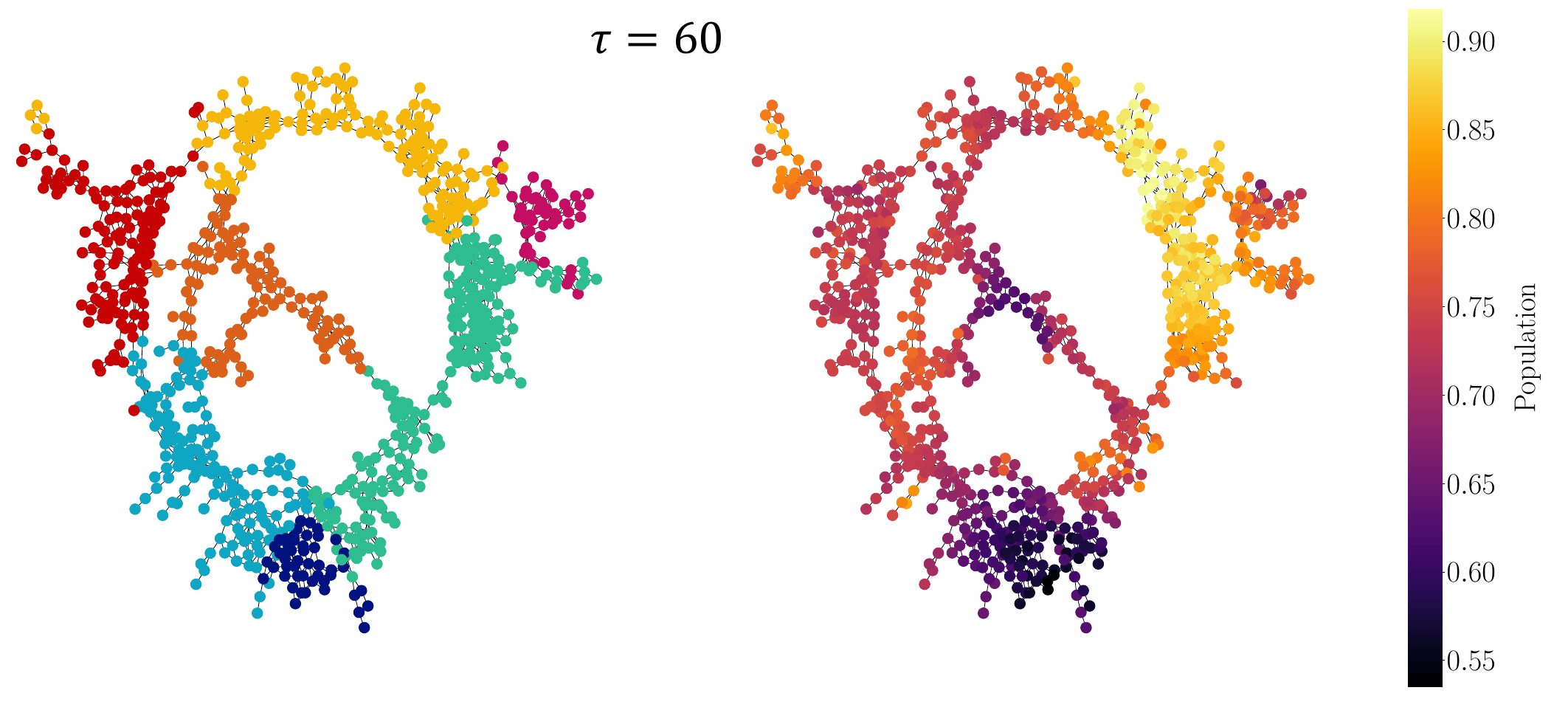}
    \includegraphics[width=\textwidth]{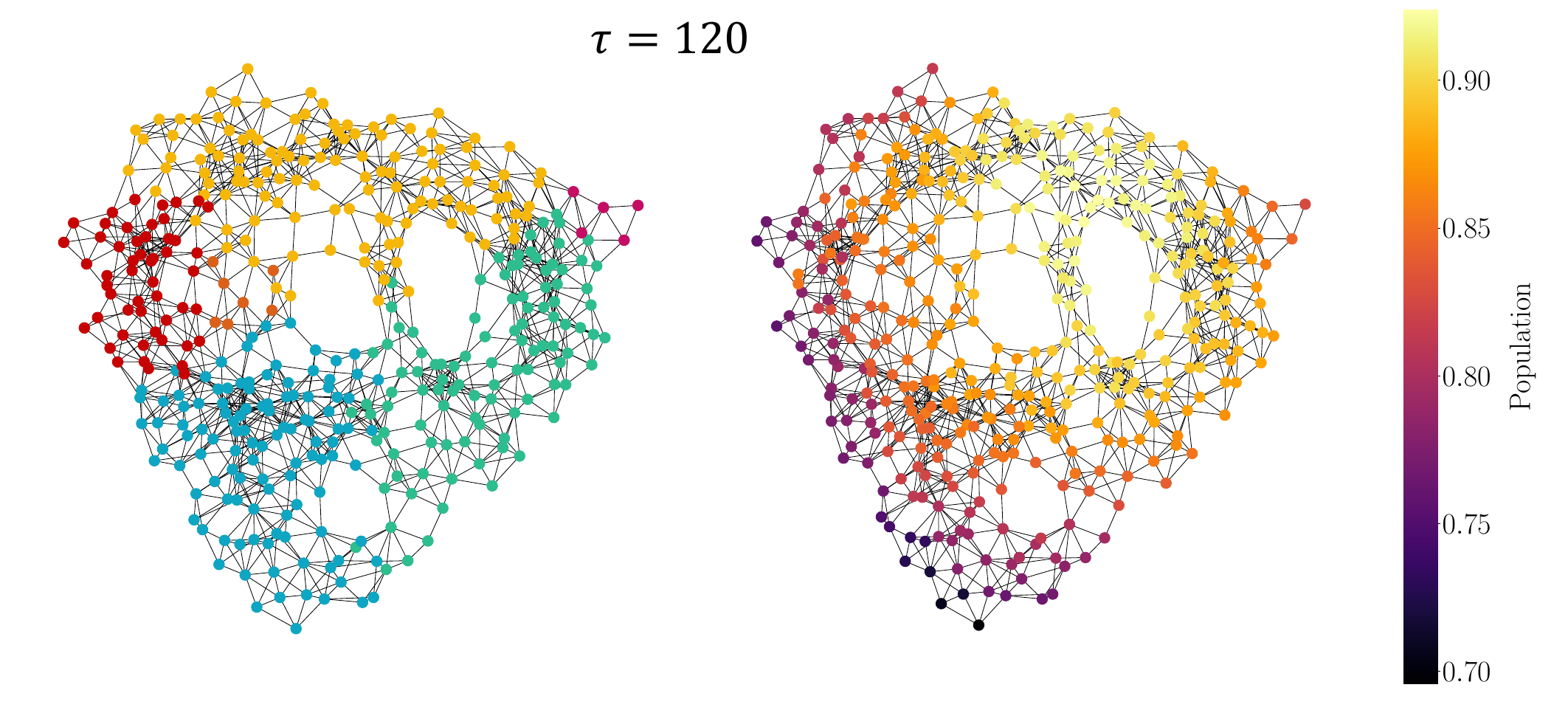}
    \includegraphics[width=\textwidth]{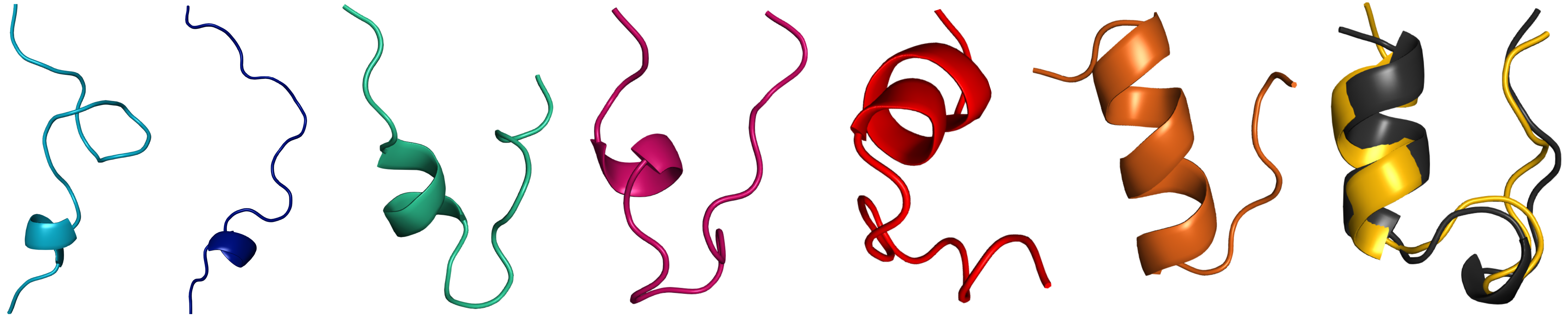}
\caption{EBC results for the folding trajectory of the Trp-cage mini-protein using topological cluster assignment with $\tau$~=~60 (top) and $\tau$~=~120 (middle): (Left): Spectral cluster assignment indicated by the node color. (Right): Population density from low (black) to high (yellow). (Bottom): Most populated state of each cluster shown as a representative structure, colored by the respective cluster color. 
The representative structure of the most populated cluster (yellow) is shown as an overlay over the experimentally determined structure (black).}\label{fig:trp_clusters}
\end{figure}

\subsection{Folding of Trp-cage Mini-protein}\label{sec:trp_cage}
As the last example, the EBC algorithm was applied to a MD trajectory of the Trp-cage mini-protein.  
Protein folding presents an interesting test case for EBC due to the connection between folding and the underlying free-energy landscape, which is widely used to conceptualize protein folding \cite{FoldingLandscape1, FoldingImplicit2, FoldingImplicit3, FoldingLandscape4}.
Trp-cage mini-protein was demonstrated to fold in implicit solvation models within accessible time scales (several hundred ns to a few $\mu$s) \cite{FoldingImplicit1, FoldingImplicit2, FoldingImplicit3}.
To visualise the temporal dynamics, the procedure for topological cluster assignment (see Section \ref{sec:topological_clustering}) was used with $\tau$~=~$60$ and $120$. The resulting graphs are shown in Figure \ref{fig:trp_clusters}a and b.

As expected, the graph shrinks for the larger $\tau$ value (Figure \ref{fig:trp_clusters}b). This trend is explained by the fact that at longer time scales the dynamics will be governed by the same few metastable states. Mapping the population density onto the nodes, we further find aggregation in certain regions (coloring in the right panels of Figure \ref{fig:trp_clusters}). 
Extracting $n=7$ clusters with spectral cluster assignment ($\tau=60$), we find that the most highly populated regions in Figure \ref{fig:trp_clusters} correspond to the structure (yellow) with the lowest RMSE of 1.4\AA with respect to the experimentally determined reference structure.
An overlay of the representative structure with the experimental reference structure (black) is shown at the bottom of Figure \ref{fig:trp_clusters}. Similar observations can be made for $\tau=120$ (middle panels in Figure \ref{fig:trp_clusters}).
Interestingly, the graph visualisations suggest the existence of two routes towards the folded state, which agrees qualitatively with a previous study of the Trp-cage folding pathway \cite{TRPFoldingPathway}.
Furthermore, we observe how certain unfolded structures at the shorter time frame (e.g., the dark blue cluster in the top left panel of Figure \ref{fig:trp_clusters}) vanish for $\tau=120$, showing how dynamics on different time scales will be dominated by different conformational states.

\section{Conclusion}
We proposed a novel clustering algorithm, which makes use of information about the potential-energy surface instead of the density of the sampled data points. 
The EBC algorithm was shown to perform robustly for difficult settings such as insufficiently sampled data or data where the sampling distribution differs from the underlying distribution.

Including information about the potential energy may be particularly interesting for systems where a potential energy (or likelihood) function is directly available, for instance for the simulation of physical systems with MD or evolutionary processes.
However, performance gains compared to other algorithms may be observed in more general settings if data points are highly concentrated thanks to the subsampling afforded by substitution of the sampling density with energies.

\section*{Acknowledgment}
The authors thank Candide Champion and Jessica Braun for helpful discussions.
This research was supported by the NCCR MARVEL, a National Centre of Competence in Research, funded by the Swiss National Science Foundation (grant number 182892).

\section*{Data and Software Availability}
An implementation of the EBC algorithm including several examples is available at \url{https://github.com/rinikerlab/EnergyBasedClustering}

\newpage

\section*{Appendix}
\subsection*{Test Systems from Scikit-learn}~\label{sec:sklearn}
\begin{figure}[H]
    \centering
    \includegraphics[width=0.3\textwidth]{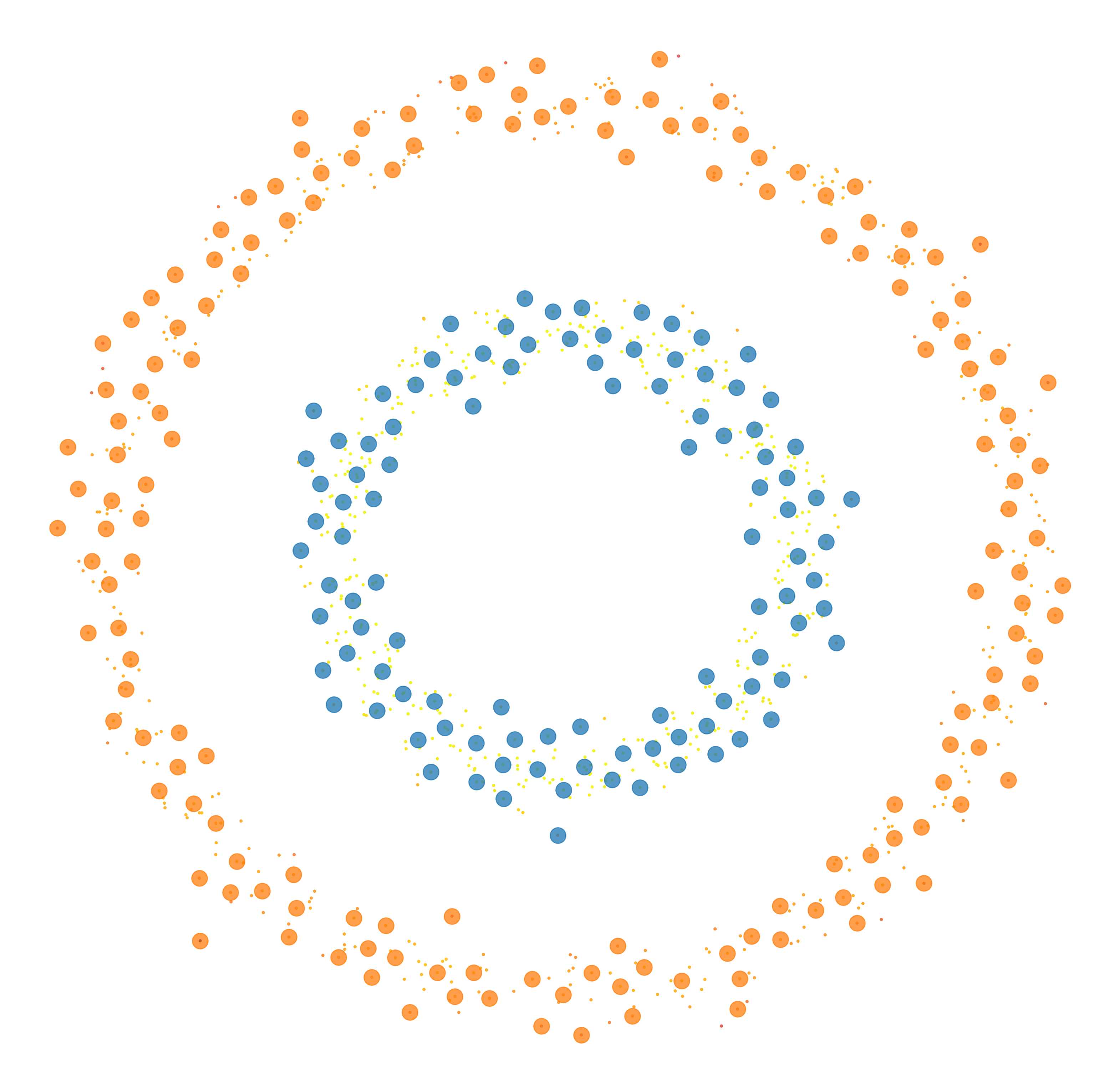}
    \includegraphics[width=0.3\textwidth]{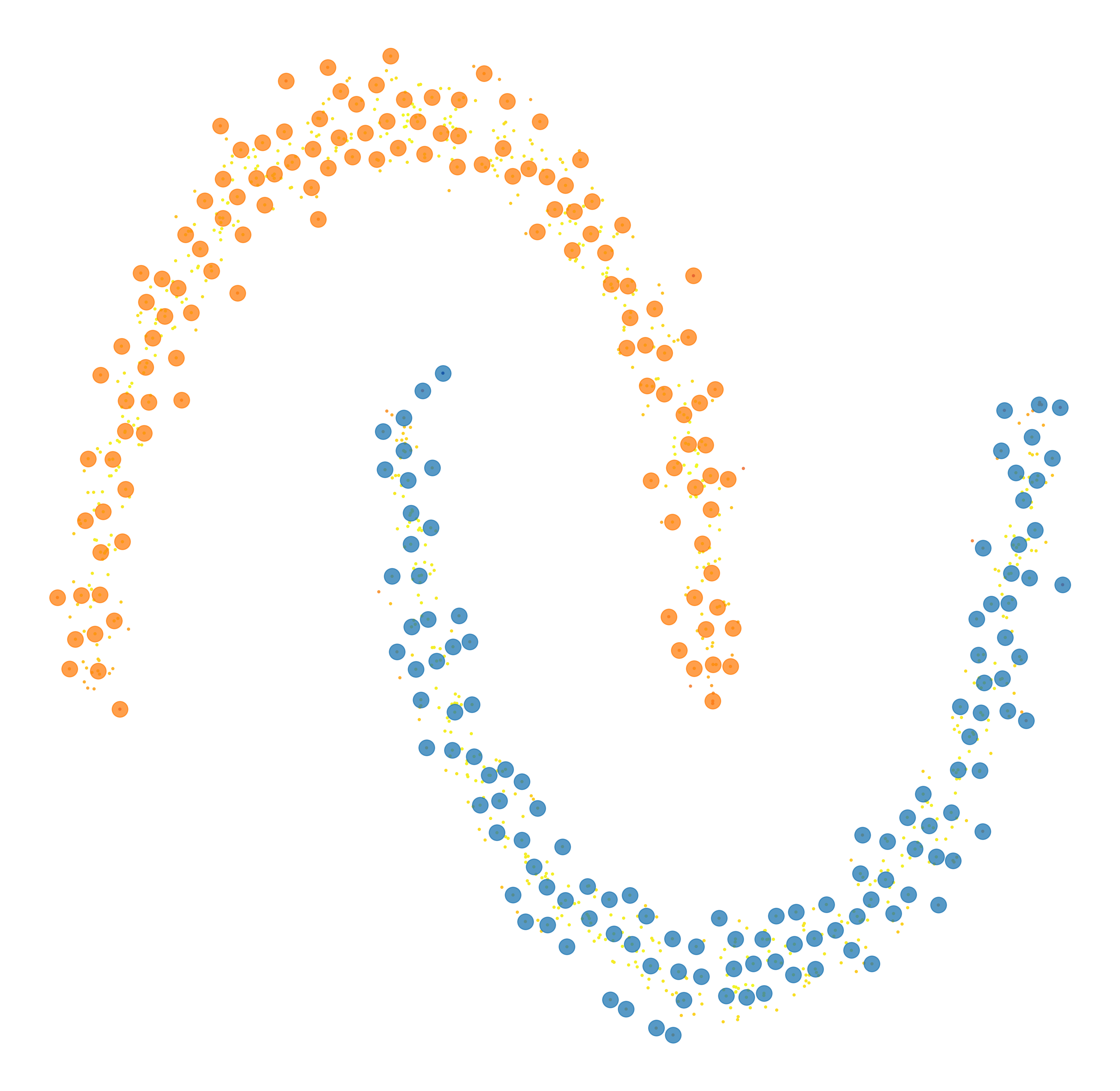}
    \includegraphics[width=0.3\textwidth]{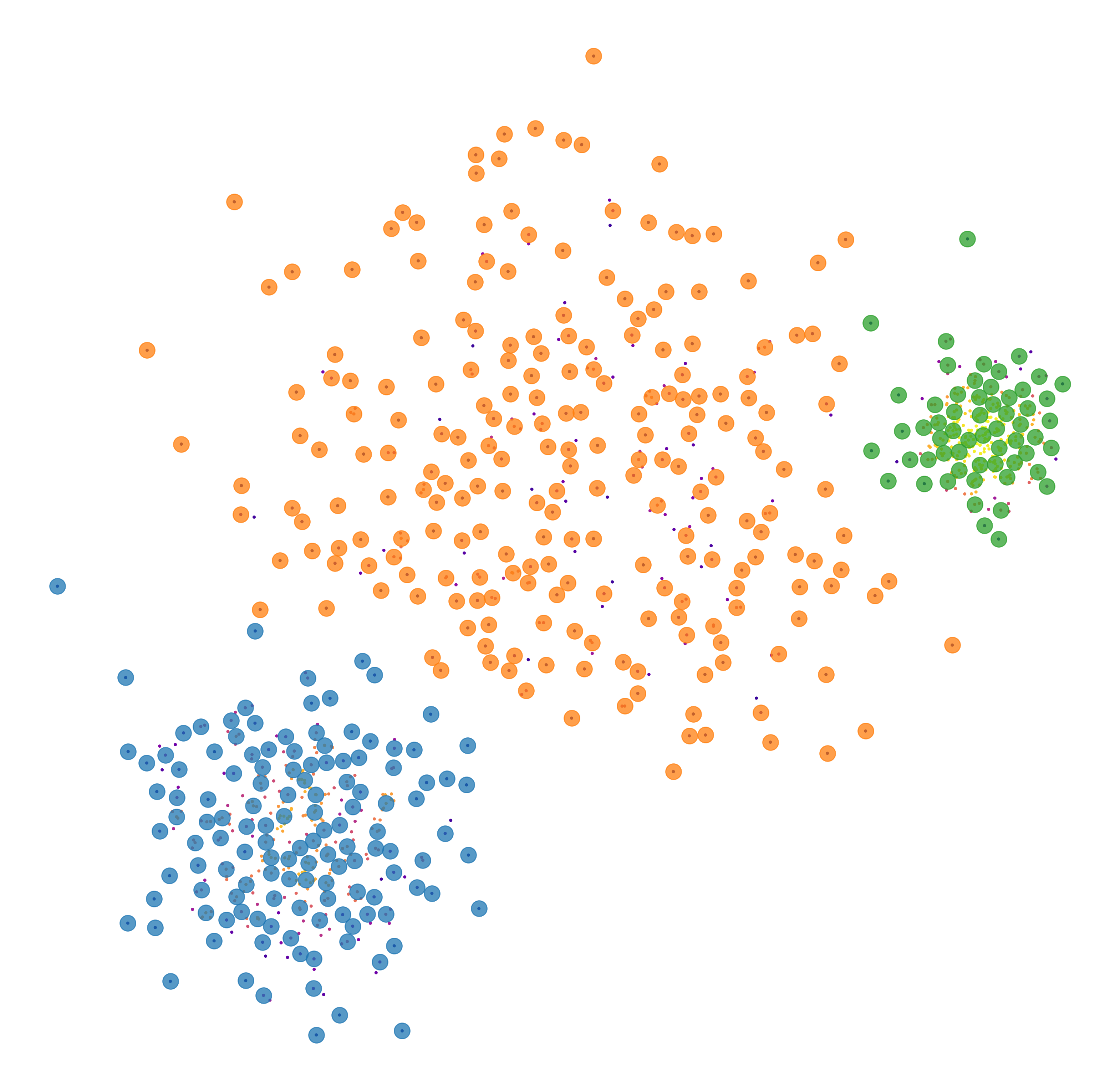} \\
    \includegraphics[width=0.3\textwidth]{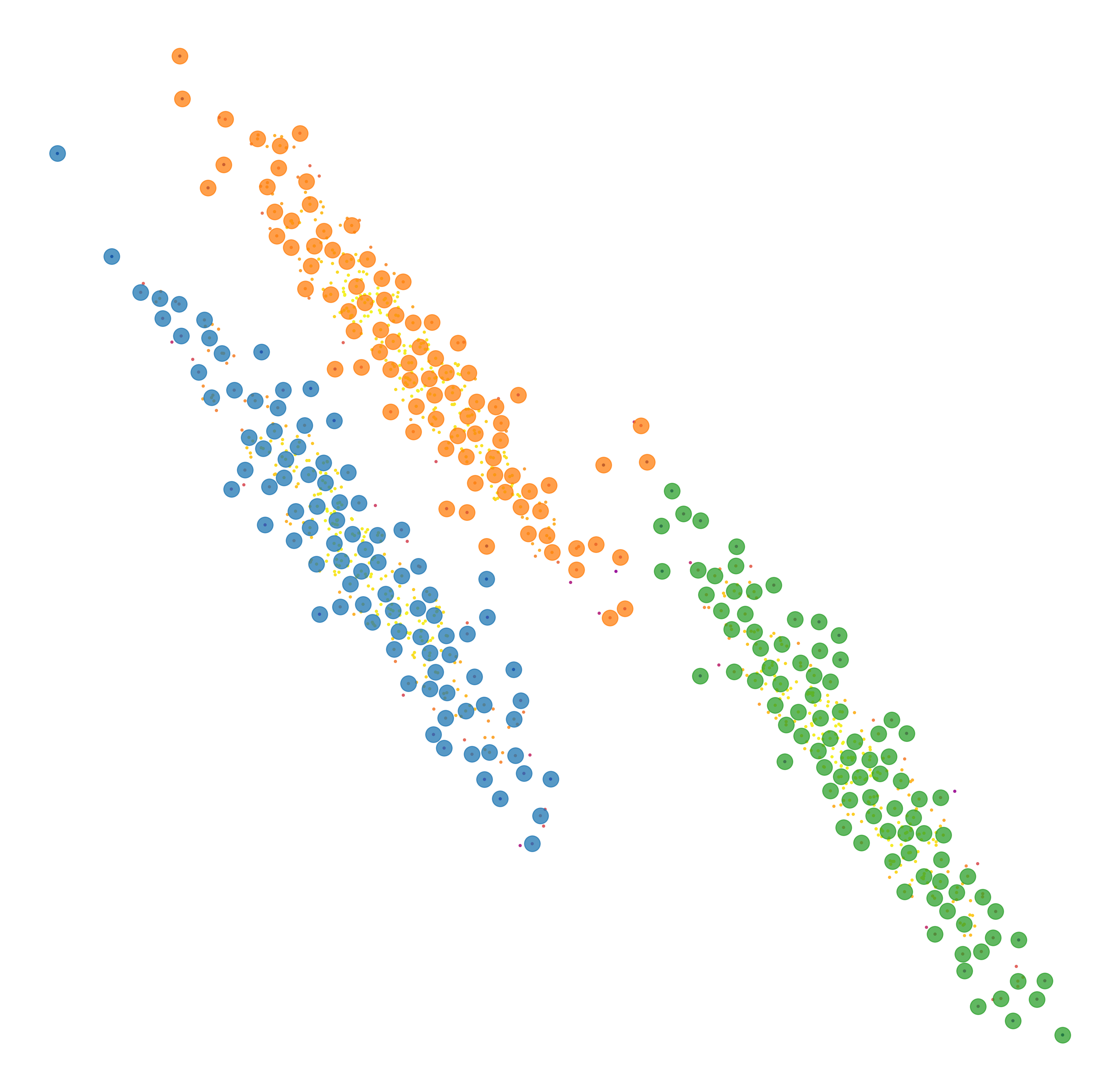}
    \includegraphics[width=0.3\textwidth]{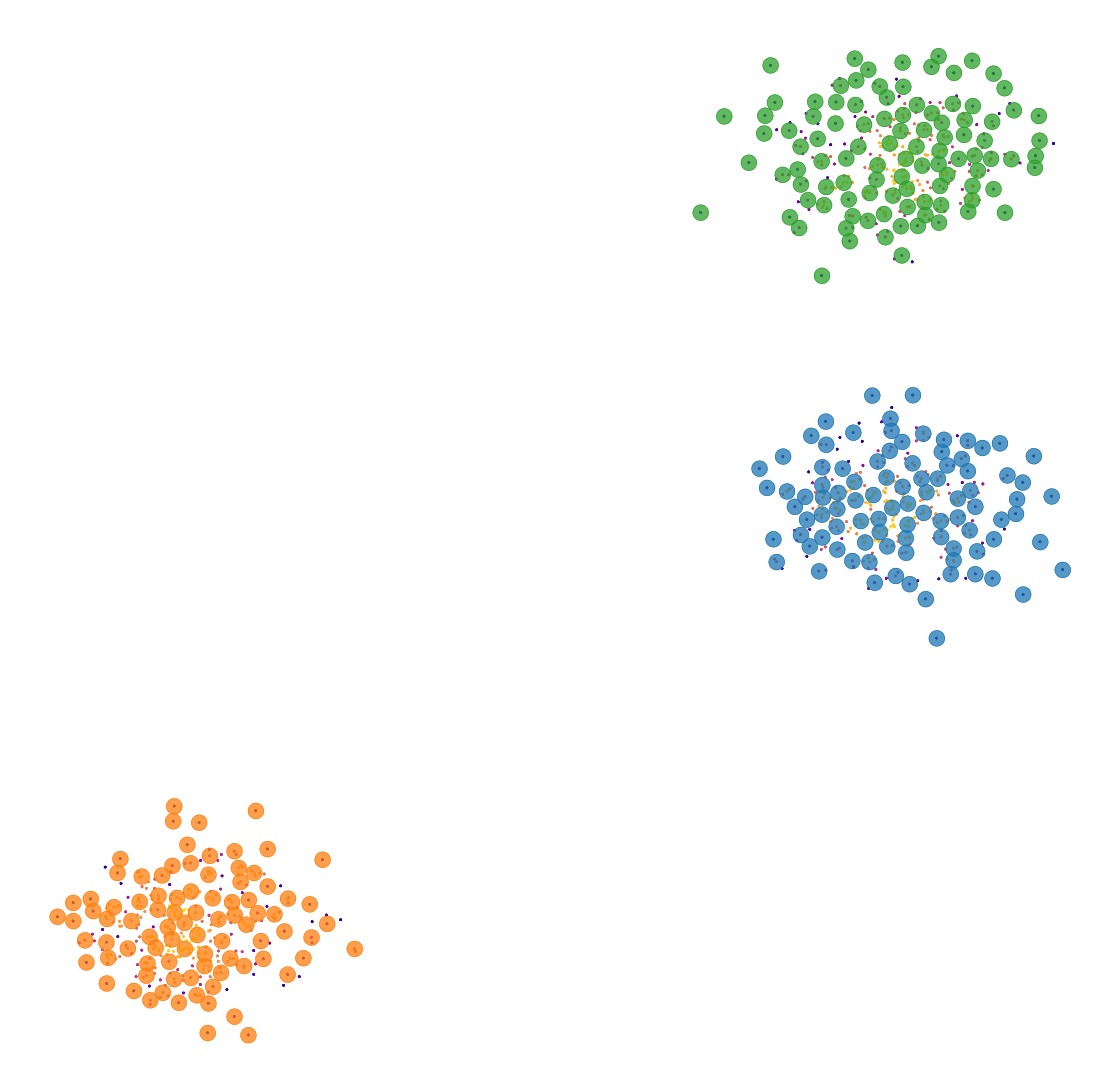}
    \includegraphics[width=0.3\textwidth]{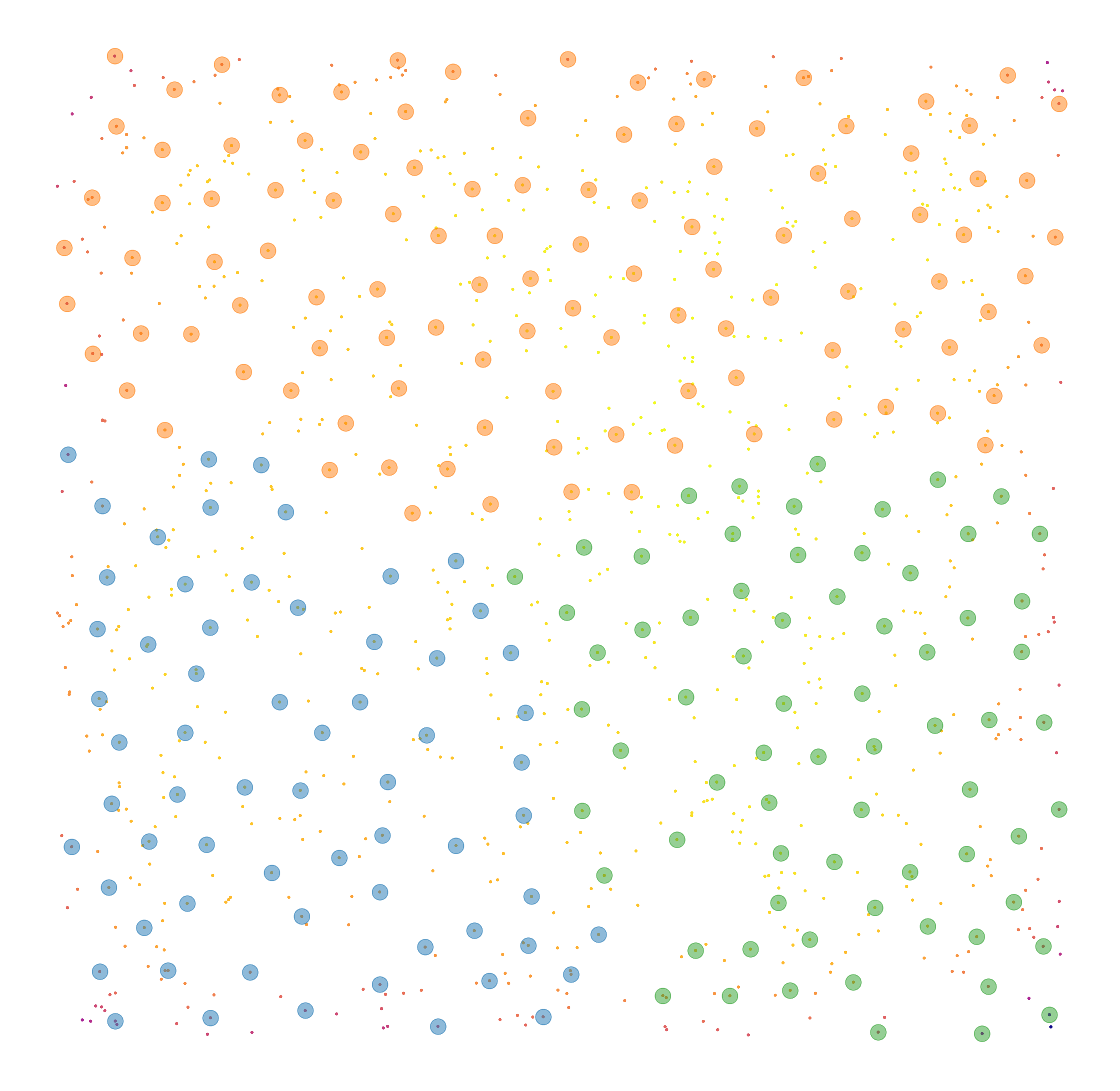}
\caption{EBC results for the standard test systems in the scikit-learn package.\cite{SKLearn}}
\label{fig:appendix_sklearn}
\end{figure}

For comparison with other clustering algorithm, results with the EBC algorithm are shown in Figure \ref{fig:appendix_sklearn} for the standard test systems in the scikit-learn package \cite{SKLearn}. The energy was obtained as the negative exponential of the log-likelihood. The log-likelihood was estimated using the Gaussian kernel density estimation implemented in scikit-learn \cite{SKLearn} using a bandwidth of $0.1$.
For each dataset, $1'000$ points were sampled.
Clusters were extracted using the the spectral cluster assignment.\cite{QRRank} 

\newpage

\printbibliography

\end{document}